\documentclass[aps,pra,reprint,groupedaddress,showkeys]{revtex4-2}
\usepackage{amsmath,graphicx,hyperref,verbatimbox,array,float,bm}
\usepackage{indentfirst}
\usepackage{braket}
\usepackage{diagbox}
\usepackage{amssymb}
\usepackage{bbold}
\usepackage{colortbl}
\usepackage{multirow}
\usepackage{subcaption}
\newcommand{\figref}[1]{Fig.~\ref{#1}}
\newcommand{\tabref}[1]{Tab.~\ref{#1}}
\renewcommand{\eqref}[1]{Eq.~\ref{#1}}

\newcommand{\centered}[1]{\begin{tabular}{l} #1 \end{tabular}}

\begin{document}

\title{Hamiltonian tomography by the quantum quench protocol with random noise}
\author{Artur Czerwinski}
\email{aczerwin@umk.pl}
\affiliation{Institute of Physics, Faculty of Physics, Astronomy and Informatics \\ Nicolaus Copernicus University,
Grudziadzka 5, 87--100 Torun, Poland}

\begin{abstract}
In this article, we introduce a framework for Hamiltonian tomography of multi-qubit systems with random noise. We adopt the quantum quench protocol to reconstruct a many-body Hamiltonian by local measurements that are distorted by random unitary operators and time uncertainty. In particular, we consider a transverse field Ising Hamiltonians describing interactions of two spins $1/2$ and three-qubit Hamiltonians of a heteronuclear system within the radio-frequency field. For a sample of random Hamiltonians, we report the fidelity of reconstruction versus the amount of noise quantified by two parameters. Furthermore, we discuss the correlation between the accuracy of Hamiltonian tomography and the number of pairs of quantum states involved in the framework. The results provide valuable insight into the robustness of the protocol against random noise.
\end{abstract}
\maketitle

\section{Introduction}

The advent of quantum technologies triggered the need for precise characterization of quantum resources. For any quantum object, the problem of its reconstruction from measurements is called quantum tomography. Traditionally, three types of quantum tomography are distinguished. Quantum state tomography (QST) aims at reconstructing the accurate quantum state of a physical system \cite{James2001,paris04,Qi2017}. Then, there is quantum process tomography (QPT) with the goal of obtaining a description of evolution in terms of a dynamical map \cite{Altepeter2003,OBrien2004,Mohseni2008}. Finally, we have quantum measurement tomography (or detector tomography) that allows one to determine the characteristics of the actual measurement operators \cite{Luis1999,Lundeen2009,Consul2020}.

A major difficulty related to quantum tomography concerns the amount of resource required for the accurate determination of a quantum object. Thus, economic frameworks that aim at optimal recovering of mathematical representations of quantum objects are gaining in popularity \cite{Gross2010,Flammia2012,Petz2012,Martinez2019,Quek2021}. In particular, in QST, one can perform complete state reconstruction with one measurement operator, which was demonstrated experimentally \cite{Oren2017}.

Various quantum phenomena are generated by many-body Hamiltonians. Thus, we need to explore characterizing techniques to estimate the parameters of Hamiltonians, which is referred to as quantum Hamiltonian tomography (QHT) \cite{Schirmer2004,Franco2009,Granade2012,Gentile2021}. There are general methods based on QST and QPT that are resource-demanding and challenging in large quantum systems. Therefore, this line of research requires a separate approach, although it lies within the scope of QPT. As a result, efficient schemes have been proposed to reduce the complexity and the need for resource in QHT. For example, a protocol involving time-dependent measurement records was theoretically proposed for Hamiltonian identification \cite{Zhang2014}. Later, it was realized experimentally on a liquid nuclear magnetic resonance quantum information processor \cite{Hou2017}. Other approaches involve estimation of quantum Hamiltonians via compressive sensing \cite{Shabani2011} or by local measurements \cite{Bairey2019}. Machine learning methods were also proposed and experimentally tested to estimate a Hamiltonian from local measurements on its ground states \cite{Xin2019}.

Recently, a new method for QHT has been proposed \cite{Li2020}. The protocol relies on utilizing multiple pairs of generic initial and final quantum states related by evolving with the Hamiltonian for an arbitrary time interval. The framework requires only local measurements on the initial and final states. The method has been tested both numerically and experimentally \cite{Zhao2021,Chen2021}. Since it involves many pairs of states with a fixed quench duration, the protocol is immune to the complexity in experiments, and the Hamiltonian can be determined uniquely. In spite of its advantages, the protocol is sensitive to other sources of experimental errors since physical models suffer from intrinsic limitations as actual measurement operators cannot be known precisely. Therefore, in the present article, we investigate the vulnerability of the quench protocol to random noise that is introduced by means of unitary operators and time uncertainty.

Throughout the article, we assume to operate in a finite-dimensional Hilbert space with the standard basis. Quantum states (initial and final) of a physical system are represented by vectors from the Hilbert space. In Sec.~\ref{methods}, we first revise the quantum quench protocol for QHT. Then, we incorporate the errors into the scheme. Next, we discuss how one can optimally estimate the Hamiltonian based on the measurable data. We also introduce a figure of merit to quantify the performance of the framework. In Sec.~\ref{results1}, the findings for one-qubit systems are presented and discussed. For three types of Hamiltonians, we utilize the framework with different numbers of pairs of initial and final states. The results allow one to observe how the quality of QHT depends on the amount of experimental noise. Then, in Sec.~\ref{results2} and \ref{results3}, we introduce the results for two-qubit and three-qubit systems, respectively. The article is concluded with a discussion and an outline of future research.

\section{Framework for Hamiltonian tomography with random noise}\label{methods}

Following Ref.~\cite{Li2020}, we implement a practical protocol for QHT, using pairs of initial and final states related by time evolution. First, we assume that the time-independent Hamiltonian of a physical system can be decomposed as:
\begin{equation}\label{f1}
\mathcal{H} = \sum_{j=1}^{\eta} \alpha_j \,\mathcal{M}_j,
\end{equation}
where $\{ \mathcal{M}_j\}$ are Hermitian matrices and $\{\alpha_j\}$ is a set of parameters characterizing the Hamiltonian. The parameters are assumed to be unknown. Then, Hamiltonian tomography involves determining the parameters $\{\alpha_j\}$ based on data that is experimentally accessible.

The framework for QHT is based on utilizing pairs of quantum systems such that one of them is prepared in the initial state $\ket{\psi_k (0)}$, whereas the other undergoes unitary evolution for a fixed quench time $T$ towards the final state $\ket{\psi_k (T)}$. These states are connected by time evolution generated by the Hamiltonian \cite{Nielsen2000}:
\begin{equation}\label{f2}
\ket{\psi_k (T)} =  e^{- i\, \mathcal{H}\, T} \ket{\psi_k (0)}
\end{equation}
for $k=1,\dots, r$, where $r$ denotes the number of pairs involved in the scheme. The states $\ket{\psi_k (0)}$ and $\ket{\psi_k (T)}$ are also linked from the perspective of energy conservation:
\begin{equation}\label{f3}
\bra{\psi_k (0)} \mathcal{H} \ket{\psi_k (0)} = \bra{\psi_k (T)} \mathcal{H} \ket{\psi_k (T)}.
\end{equation}
If we substitute \eqref{f1} into \eqref{f3}, we obtain a matrix equation: $\pmb{P} \pmb{\alpha} = \pmb{0}$, where $\pmb{P}$ is a $r\times \eta$ matrix with entries:
\begin{equation}\label{f4}
p_{kj}= \bra{\psi_k (0)} \mathcal{M}_j \ket{\psi_k (0)} - \bra{\psi_k (T)} \mathcal{M}_j \ket{\psi_k (T)}
\end{equation}
and $\pmb{\alpha}$ denotes the coefficient vector: $\pmb{\alpha} = [\alpha_1, \alpha_2, \dots, \alpha_{\eta}]^T$. Assuming that the matrices $\mathcal{M}_j$ represent measurable observables, we notice that the matrix $\pmb{P}$ is computable on the basis of an experiment. Naturally, we expect that our source can repeatedly prepare systems in one of the initial states $\ket{\psi_k (0)}$, which means that each physical copy is measured only once, and we neglect the post-measurement state of the system.

In theory, $r= \eta -1$ is sufficient to determine $\pmb{\alpha}$ up to an overall multiplicative factor provided neither of $\ket{\psi_k (0)}$ is an eigenstate of $\mathcal{H}$. In addition, larger $T$ is preferable, which implies that the initial and final states are more distinguishable. Also, initial states in the ensemble should be sufficiently distinct to provide independent information about the Hamiltonian \cite{Li2020}.

In practice, we should bear in mind that noise and errors are inherent to any kind of measurement. Therefore, it appears justifiable to perform the procedure for a higher number of pairs to collect sufficient information for a precise estimation of the Hamiltonian.

One specific type of error that may occur in this scheme relates to inaccuracies in the settings of the measurement setup. Depending on the nature of the physical system, a measurement operator $\mathcal{M}_j$ can be experimentally realized in different ways. In any case, due to settings uncertainty, the act of measurement does not correspond strictly to $\mathcal{M}_j$ but to a distorted operator denoted by $\widetilde{\mathcal{M}}_j$. The distortion can be mathematically modeled by random unitary matrices that act on the original measurement operator \cite{Lohani2020,Danaci2021}:
\begin{equation}\label{f6}
U (\omega_1, \omega_2, \omega_3) = \begin{pmatrix} e^{ - i/2 (\omega_1+ \omega_3)} \cos \frac{\omega_2}{2} & - e^{ i/2 (\omega_3 - \omega_1)} \sin \frac{\omega_2}{2}  \\\\  e^{ i/2 (\omega_1 - \omega_3)} \sin \frac{\omega_2}{2}  & e^{ i/2 (\omega_1 + \omega_3)} \cos \frac{\omega_2}{2} \end{pmatrix},
\end{equation}
where $\omega_1, \omega_2, \omega_3$, in our application, are selected randomly from a normal distribution characterized by the mean value equal $0$ and a non-zero standard deviation denoted by $\sigma$, i.e. $\omega_1, \omega_2, \omega_3 \in \mathcal{N}(0,\sigma)$. Assuming that the Hamiltonian $\mathcal{H}$ describes $N-$qubit systems, for every measurement operator $\mathcal{M}_j$, we construct a perturbation matrix as:
\begin{equation}\label{f61}
Q_j (\sigma) = U (\omega_1^1, \omega_2^1, \omega_3^1)  \otimes \dots \otimes U (\omega_1^N, \omega_2^N, \omega_3^N),
\end{equation}
which allows us to compute a distorted measurement operator:
\begin{equation}\label{f5}
\widetilde{\mathcal{M}}_j (\sigma) := Q_j (\sigma)  \,\mathcal{M}_j\, Q_j^{\dagger} (\sigma).
\end{equation}
With a given $\sigma$, a different perturbation matrix $Q_j (\sigma) $ is generated for each measurement operator $\mathcal{M}_j$. Thanks to this approach, every measurement is burdened with a random error, and $\sigma$ is used to quantify the amount of experimental noise. Then, the matrix defined in \eqref{f4} depends on $\sigma$ and can be denoted as $\pmb{P} (\sigma)$. When the linear equation $\pmb{P} (\sigma) \pmb{\alpha} = \pmb{0}$, it is solved by Lagrange multipliers as follows \cite{Zhao2021}:
\begin{equation}\label{f7}
\begin{split}
&\pmb{P}^T  (\sigma) \pmb{P} (\sigma) \pmb{\alpha} - \lambda \pmb{\alpha} = 0,\\
& \pmb{\alpha}^T \pmb{\alpha} = 1,
\end{split}
\end{equation}
where $\lambda$ stands for the Lagrange multiplier. To estimate the optimal $\pmb{\alpha}$ we follow the method of least squares (LS). We define the error function as $f  =  \pmb{\alpha}^T \pmb{P}^T  (\sigma) \pmb{P} (\sigma) \pmb{\alpha}$ and find the solution of the linear regression problem as the singular vector of $\pmb{P} (\sigma)$ corresponding to the minimal singular value \cite{Zhao2021}.

Apart from the unitary rotations, also time uncertainty can be imposed on the measurement results. The time interval for the evolution of the system, $T$, can be selected arbitrarily. However, when performing measurements to compute the elements of the matrix $p_{ij}$, one is exposed to the detector's timing jitter, which can be modeled by a Gaussian distribution, cf. \cite{Sedziak2020}. Therefore, every moment of observation should be selected from a normal distribution, i.e. $p_{ij} \rightarrow T_{ij} \in  \mathcal{N}(T,\Delta \tau)$, where $\Delta \tau$ denotes the standard deviation, and is used to quantify the amount of time uncertainty. When the extended noise scenario is considered, we obtain a matrix burdened with two kinds of error: $\pmb{P} (\sigma, \Delta \tau)$, which can be used to estimate the Hamiltonian according to the LS method described above.

We introduce a figure of merit to quantify the accuracy of QHT with random noise. If $\widetilde{\pmb{\alpha}}$ stands for the result of estimation, we compute fidelity:
\begin{equation}\label{f8}
\mathcal{F} := |\cos \sphericalangle (\pmb{\alpha}, \widetilde{\pmb{\alpha}})|,
\end{equation}
which allows us to compare the outcome of the framework with the actual vector for a given Hamiltonian. To evaluate the average performance of the framework, we utilize a sample of Hamiltonians \eqref{f1}. For each input Hamiltonian, we generate noisy data and perform its reconstruction according to the quantum quench protocol. Then we compute the mean value of the fidelity \eqref{f8} along with the sample standard deviation (SD). Depending on the parameter selected as independent variable, we treat the average fidelity either as a function of $\sigma$ or $\Delta \tau$, which is denoted by $\mathcal{F}_{av} (\sigma)$ and $\mathcal{F}_{av} (\Delta \tau)$, respectively.

The scheme outlined in this section can be implemented numerically to investigate the efficiency of QHT for different measurement bases. We first consider three types of one-qubit Hamiltonians decomposed by means of different sets of operators. Next, we investigate two-qubit and three-qubit Hamiltonians. In each case, the initial states were selected arbitrarily so that they were sufficiently distinct with respect to the geometric distance. The ensemble of initial states is fixed for each type of Hamiltonian. However, the exact number of states introduced into the framework varies. Consequently, we can examine how the number of pairs influences the accuracy of Hamiltonian reconstruction with random noise. Furthermore, taking into account the noise introduced by random rotations and time uncertainty, the results allow one to observe how the quality of QHT is affected by different parameters.

\section{Hamiltonian tomography of one-qubit systems}\label{results1}

\subsection{Hamiltonian in a basis defined by the SIC-POVM}

As the first example, let us consider QHT of two-level systems governed by:
\begin{equation}\label{r1}
\mathcal{H} = \sum_{j=1}^3 \alpha_j \ket{\xi_j} \! \bra{\xi_j},
\end{equation}
where 
\begin{equation}\label{r2}
\begin{split}
&\ket{\xi_1} = \frac{1}{\sqrt{3}} \ket{0} + \sqrt{\frac{2}{3}} \ket{1}, \\
&\ket{\xi_2} = \frac{1}{\sqrt{3}} \ket{0} + \sqrt{\frac{2}{3}} e^{i \frac{2 \pi}{3}} \ket{1},\\
&\ket{\xi_3} = \frac{1}{\sqrt{3}} \ket{0} + \sqrt{\frac{2}{3}}  e^{i \frac{4 \pi}{3}} \ket{1}
\end{split}
\end{equation}
and $\{ \ket{0}, \ket{1}\}$ denotes the standard basis in the $2-$dimensional Hilbert space. The vectors $\ket{\xi_1}, \ket{\xi_2}, \ket{\xi_3}$ along with the basis vector $\ket{0}$ define a symmetric, informationally complete, positive operator-valued measure (SIC-POVM) \cite{Renes2004,Fuchus2017}. The SIC-POVM can be considered a minimal set of measurement operators for qubit tomography \cite{Rehacek2004}.

\begin{figure}[h]
	\centering
		\centered{\includegraphics[width=0.99\columnwidth]{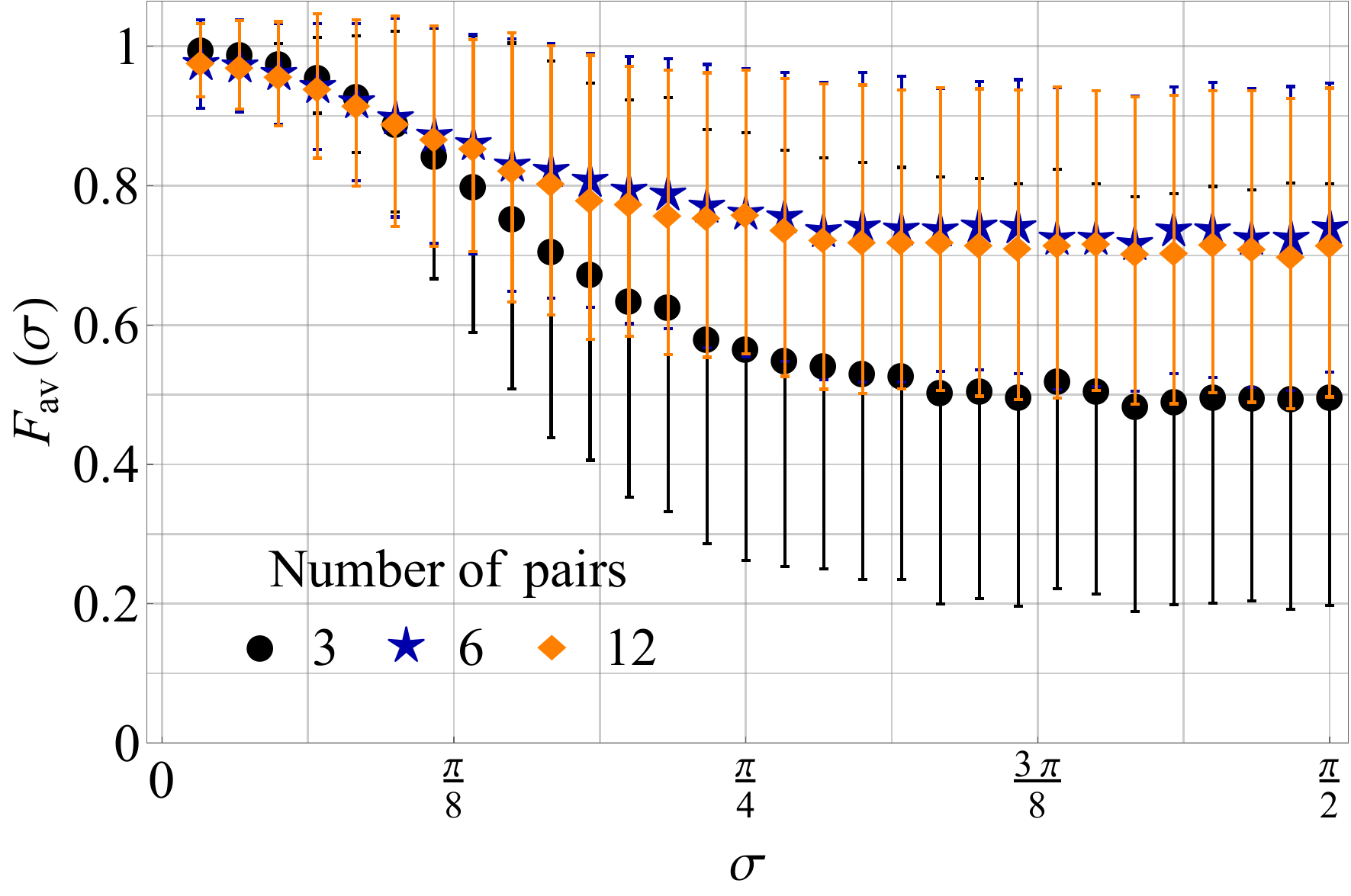}}
	\caption{Plots of $\mathcal{F}_{av} (\sigma)$ for a sample of $100$ two-qubit Hamiltonians represented by the SIC-POVM. Error bars correspond to one SD.}
	\label{hamiltonian1}
\end{figure}

Since the measurements defined by the SIC-POVM can be realized in laboratories for different physical systems, see Ref.~\cite{Bent2015,Zhao2015}, it appears relevant to investigate reconstruction of Hamiltonians decomposed in such a basis. We study QHT for a sample of $100$ random Hamiltonians of the form \eqref{r1}. In \figref{hamiltonian1}, one finds the plots of $\mathcal{F}_{av} (\sigma)$ for three numbers of pairs involved in QHT. Time uncertainty is not included, i.e., we have $\Delta \tau=0$.

First, we notice that for three pairs, one can obtain perfect accuracy for smaller values of $\sigma$. Based on numerical simulations, we can determine a threshold value for $\sigma$, which guarantees a desired quality of QHT. In particular, $\mathcal{F}_{av} (\sigma)>0.9$ as long as $\sigma < \pi/10$. Additionally, these results feature minor variation. However, when $\sigma$ is increased, $\mathcal{F}_{av} (\sigma)$ declines whereas SD grows significantly. For a greater amount of noise, the figure of merit stabilizes at $\mathcal{F}_{av} (\sigma) \approx 0.5$, but the value of SD (approximately $0.3$) implies that the results for the sample are spread out over a wide range. This means that we cannot efficiently estimate the Hamiltonian when $\sigma$ is significant.

The plot \figref{hamiltonian1} allows one to observe to what extent increasing the number of pairs influences the accuracy of QHT. We can notice that there is no significant difference between $6$ and $12$ pairs of quantum states involved in QHT. For both numbers, identical conclusions can be made. More specifically, for $\sigma < \pi/10$, we again obtain $\mathcal{F}_{av} (\sigma)>0.9$, but the samples feature more statistical dispersion than in the case of three pairs. As we increase $\sigma$, the average fidelity reduces to $\mathcal{F}_{av} (\sigma) \approx 0.71$ while the value of SD grows to $\approx 0.23$.

To conclude, one can agree that, for greater values of $\sigma$, the scenarios with more pairs ($6$ and $12$) outperforms the scheme based on three pairs in terms of both the average fidelity and standard deviations. However, even for higher numbers of pairs, the results are not sufficient to guarantee efficient QHT when the random noise is severe.

\subsection{Hamiltonian in a polarization basis}

Second example of a one-qubit Hamiltonian is defined in a basis composed of vectors that are commonly used to represent polarization states of light. Concretely, we discuss the following form:
\begin{equation}\label{r3}
\mathcal{H} = \sum_{j=1}^3 \alpha_j \ket{\psi_j} \! \bra{\psi_j},
\end{equation}
where
\begin{equation}\label{r4}
\begin{split}
&\ket{\psi_1} = \ket{0}, \hspace{1cm} \ket{\psi_2} = \frac{1}{\sqrt{2}}( \ket{0} + \ket{1}),\\
&\ket{\psi_3} = \frac{1}{\sqrt{2}} (\ket{0} + i \ket{1}).
\end{split}
\end{equation}
Conventionally, the vectors $\ket{\psi_1}, \ket{\psi_2}, \ket{\psi_3}$ represent horizontal, diagonal, and right-circular polarization states, respectively \cite{Altepeter2005}. Projective measurements defined by those vectors are commonly performed in photonic tomography. Thus, for the Hamiltonian \eqref{r3} the corresponding matrix \eqref{f4} can be considered experimentally accessible, which justifies implementation of the quantum quench protocol.

\begin{figure}[h]
	\centering
		\centered{\includegraphics[width=0.99\columnwidth]{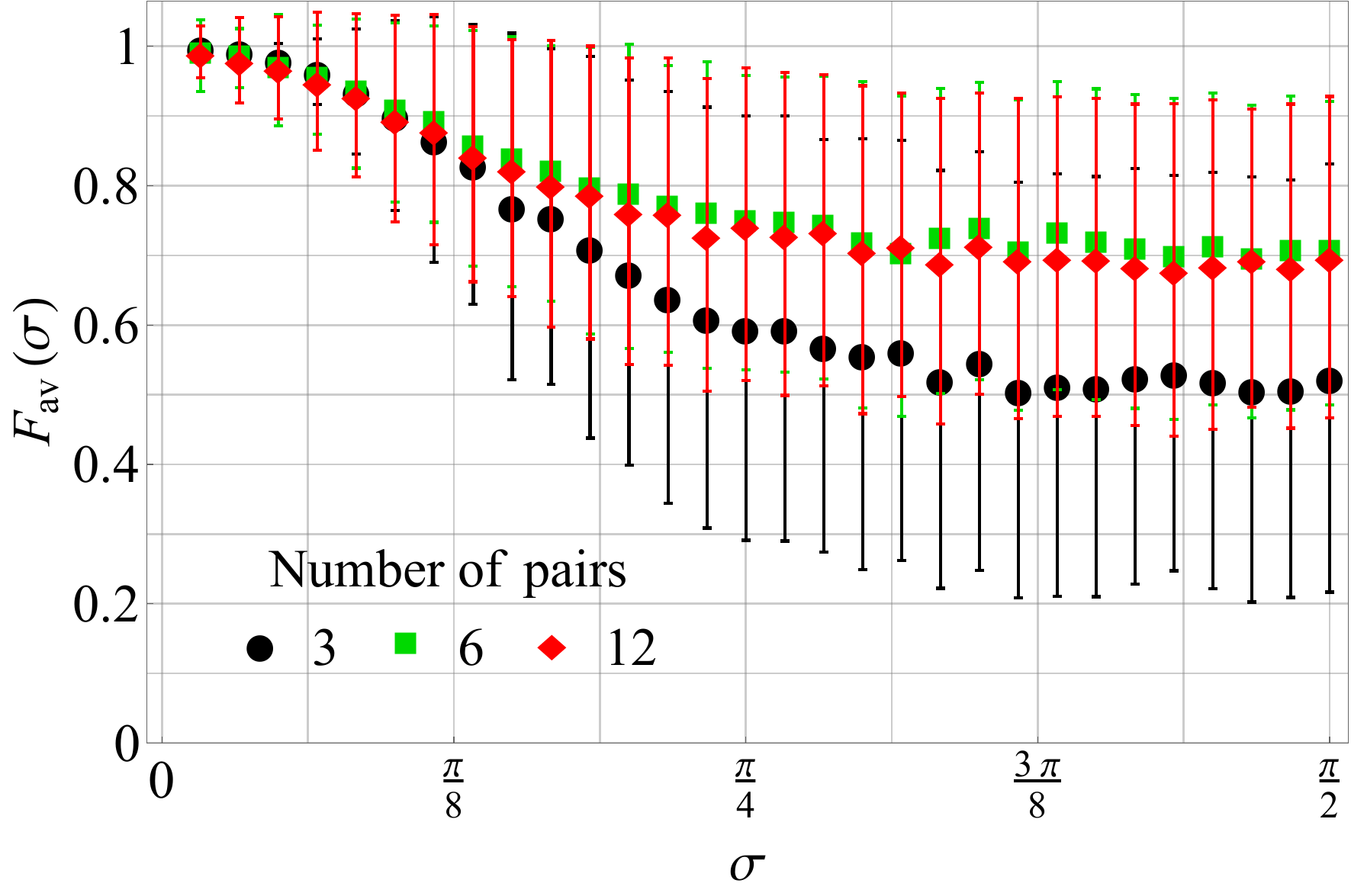}}
	\caption{Plots of $\mathcal{F}_{av} (\sigma)$ for a sample of $100$ two-qubit Hamiltonians decomposed in the polarization basis. Error bars correspond to one SD.}
	\label{hamiltonian2}
\end{figure}

In \figref{hamiltonian2}, one finds the results of numerical simulations devoted to the reconstruction of Hamiltonians decomposed in the polarization basis \eqref{r3}. Time uncertainty is not included, i.e., we have $\Delta \tau=0$. Distinct plot markers are associated with a different number of pairs involved in the framework.

We notice that the plots obey the same tendencies as in the case of the Hamiltonians decomposed in the SIC-POVM basis. For three pairs, we obtain perfect accuracy and insignificant SD as long as $\sigma < \pi/10$. For greater values of $\sigma$, the efficiency of QHT declines rapidly towards $\mathcal{F}_{av} (\sigma) \approx 0.5$ and $\mathrm{SD} \approx 0.3$. Similarly as before, increasing the number of pairs leads to greater statistical dispersion for smaller $\sigma$, but for a heavier noise, we obtain better results than with three pairs.

As to compare QHT for polarization basis with the SIC-POVM case with respect to the scenarios involving more pairs (i.e., $6$ and $12$), we observe that in the polarization basis, the SD is smaller up to $\sigma = \pi/6$, which implies that the values tend to be closer to the mean. However, the mean value itself appears to be greater for the SIC-POVM scenario, particularly when we utilize $6$ pairs of states.

\subsection{Hamiltonian in the basis of the Pauli operators}

Next, we consider a one-qubit Hamiltonian defined in the basis of the Pauli matrices:
\begin{equation}\label{r5}
\mathcal{H} = \sum_{j=1}^3 \alpha_j \sigma_j,
\end{equation}
where
\begin{equation}\label{r6}
\sigma_1 = \begin{pmatrix} 0 & 1 \\ 1 & 0  \end{pmatrix}, \hspace{0.2cm}  \sigma_2 = \begin{pmatrix} 0 & i \\ - i & 0  \end{pmatrix},\hspace{0.2cm}  \sigma_3 = \begin{pmatrix} 1 & 0\\ 0 & -1  \end{pmatrix}.
\end{equation}

\begin{figure}[h]
	\centering
		\centered{\includegraphics[width=0.99\columnwidth]{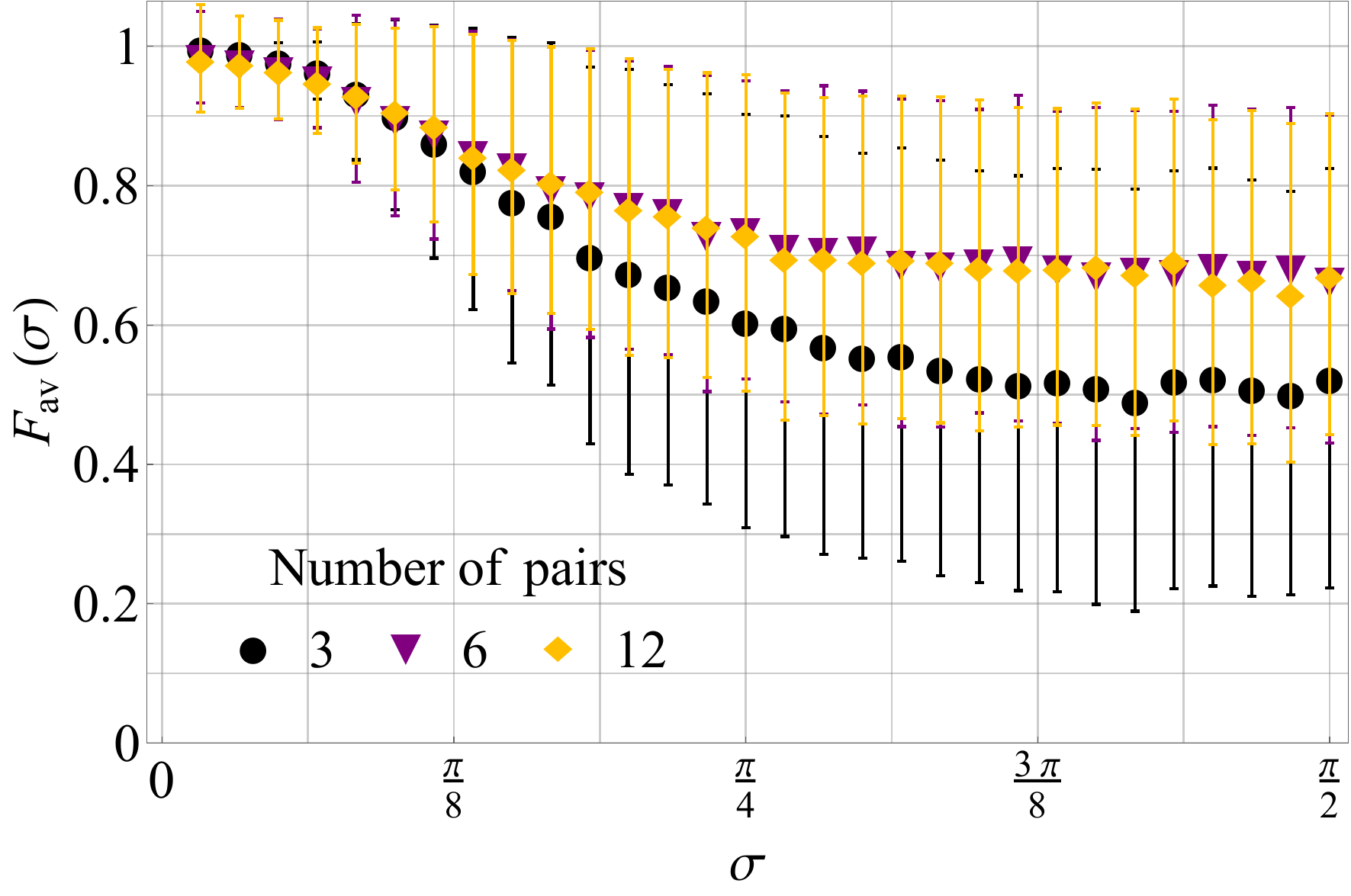}}
	\caption{Plots of $\mathcal{F}_{av} (\sigma)$ for a sample of $100$ two-qubit Hamiltonians decomposed in the Pauli basis. Error bars correspond to one SD.}
	\label{hamiltonian3}
\end{figure}

The results obtained for Hamiltonian \eqref{r5} appear very consistent with the above examples, see \figref{hamiltonian3}. We can formulate the same conclusions concerning the tendencies that can be observed in the plots. Again, we can conclude that Hamiltonian estimation is most precise with three pairs for $\sigma < \pi/10$. Furthermore, we can compare this example with the first one, and notice that for $6$ pairs of states, the framework is less robust against random noise for the Pauli basis than the SIC-POVM.

\section{Hamiltonian tomography of two-qubit systems}\label{results2}

We implement the quantum quench protocol to study estimation of two-qubit Hamiltonians within the transverse-field Ising model \cite{Stinchcombe1973,Chakrabarti1996}. It involves a lattice with nearest-neighbor interactions determined by the alignment or anti-alignment of spin projections along the $z$ axis, as well as an external magnetic field that is parallel to the $x$ axis (perpendicular to the $z$ axis). From the quantum mechanical point of view, an important feature of this model is the fact that the spin projections along the $x$ axis and the $z$ axis are not commuting observables, which implies that they cannot both be simultaneously observed. As a result, a classical approach based on statistical mechanics cannot describe this model, and a quantum treatment is needed.

\begin{figure}[h]
	\centering
		\centered{\includegraphics[width=0.95\columnwidth]{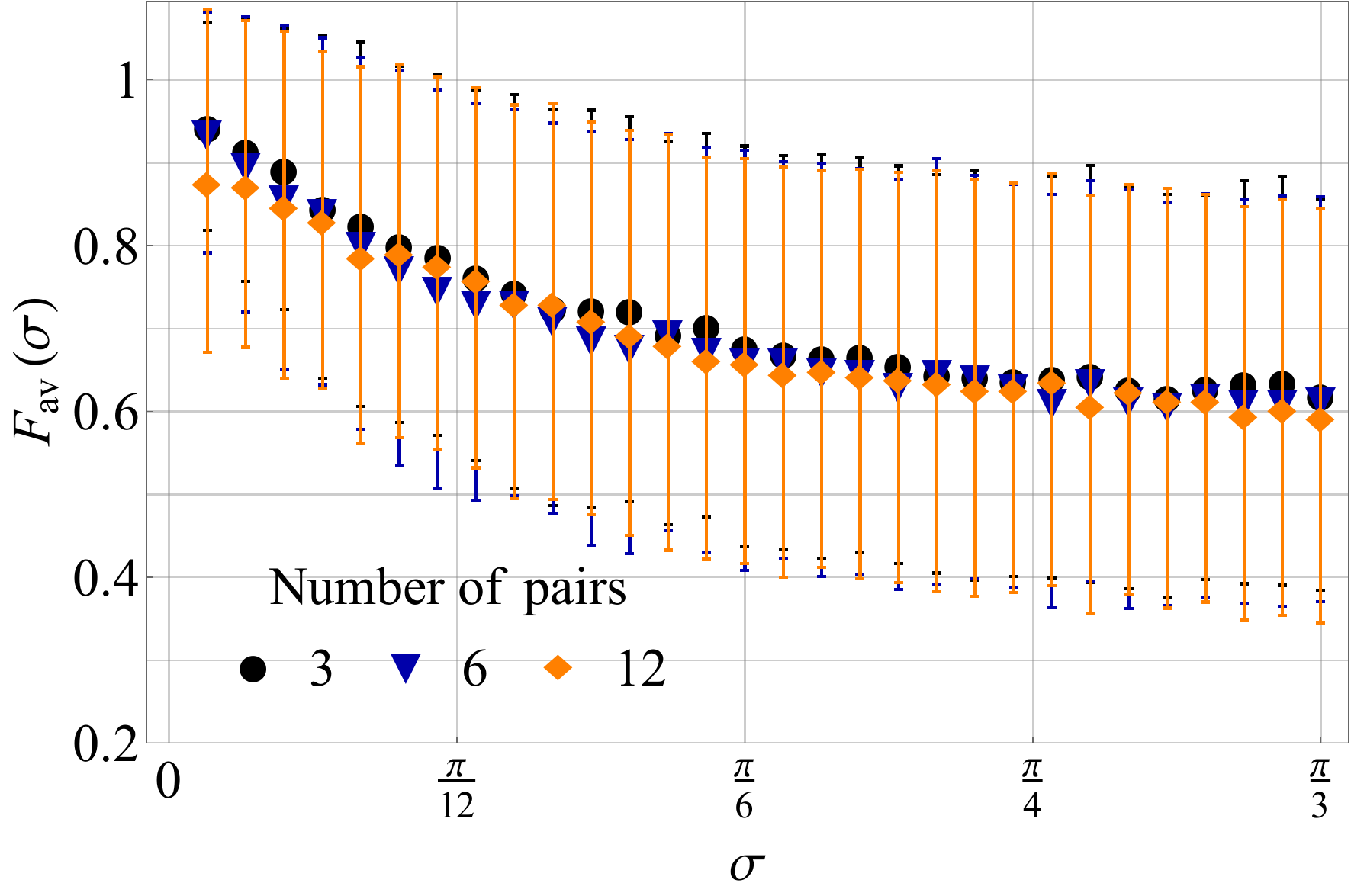}}
	\caption{Plots of $\mathcal{F}_{av} (\sigma)$ for a sample of $100$ two-qubit Hamiltonians within the transverse-field Ising model. Error bars correspond to one SD.}
	\label{hamiltonian4}
\end{figure}

To be more specific, the model has the following quantum Hamiltonian:
\begin{equation}\label{r7}
\mathcal{H}_N = \sum_{j=1}^N \alpha_j X_j + \sum_{j=1}^{N-1} \beta_j Z_j Z_{j+1},
\end{equation}
where $N$ denotes the number of qubits in the model and the symbols: $X_j$ and $Z_j$ represent the elements of the spin algebra acting on the spin variables of the corresponding elements. In the case of spin $1/2$, we implement the Pauli matrices, see \eqref{r6}.

In our simulation, we substitute $N=2$. For the sake of mathematical clarity, the two-qubit Hamiltonian can be written explicitly as:
\begin{equation}\label{r8}
\mathcal{H}_2 = \alpha_1 \,\sigma_1 \otimes \mathbb{I}_2 + \alpha_2 \,\mathbb{I}_2 \otimes \sigma_1 + \beta_1 \,\sigma_3 \otimes \sigma_3,
\end{equation}
where the coefficients $\alpha_1, \alpha_2, \beta_1$ are to be determined from measurable data.

In this case, we proceed analogously as in the above examples -- a sample of $100$ random Hamiltonians of the form \eqref{r7} is selected and reconstructed by the quantum quench protocol. Here, we utilize two-qubit perturbation matrices, see \eqref{f61}, which deform the measurement operators appearing in the Hamiltonian decomposition. First, time uncertainty is not included, which means that $\Delta \tau = 0$.

The results obtained for the two-qubit Hamiltonians are presented in \figref{hamiltonian4}. One can observe that, for any number of pairs, the quality of Hamiltonian tomography deteriorates as we increase the noise ratio. The three plots converge for greater amounts of noise, and, ultimately, we obtain $\mathcal{F}_{av} \approx 0.6$ with $SD\approx 0.25$. The properties of the plots are peculiar when minor values of $\sigma$ are considered. For $\sigma = \pi/90$, it turns out that the quality of Hamiltonian estimation is the poorest if we utilize $12$ pairs of initial and final states. In such a case, we obtain $\mathcal{F}_{av} = 0.88$ and $SD = 0.21$, whereas for the other two numbers of pairs, we obtain better figures, as presented in \tabref{table1}. Both scenarios result in $\mathcal{F}_{av} = 0.94$, but they differ in SD, which is greater in the case of $6$ pairs. The difference in statistical dispersion can be attributed to the errors that arise in numerical algorithms which estimate the Hamiltonian.

\begin{table}[h]
\setlength{\tabcolsep}{7.5pt} 
\renewcommand{\arraystretch}{2.45}
	\begin{tabular}{|c|c|c|c|c|c|c|}
\hline
\multirow{2}{*}{
				\backslashbox[9 mm]{$\sigma$}{$r$}} & \multicolumn{2}{c|}{$3$} & \multicolumn{2}{c|}{$6$} & \multicolumn{2}{c|}{$12$}\\ \cline{2-7} 
	 	&  $\mathcal{F}_{av}$ & SD & $\mathcal{F}_{av}$ & SD & $\mathcal{F}_{av}$ & SD \\ \hline
$\frac{\pi}{90}$ &$0.94$ & $0.13$  &$0.94$ & $0.16$ & $0.88$ & $0.21$   \\ \hline
$\frac{\pi}{45}$ &$0.92$ & $0.16$ &$0.90$ & $0.18$ & $0.87$ & $0.20$  \\ \hline
$\frac{\pi}{30}$ &$0.85$ & $0.21$ &$0.86$ & $0.21$ & $0.85$ & $0.21$  \\ \hline
	\end{tabular}
	\caption{Average fidelity $\mathcal{F}_{av}$ and SD for a sample of $100$ two-qubit Hamiltonians reconstructed according to the quantum quench protocol with random noise.}
	\label{table1}
\end{table}

For a sample of random Hamiltonians, it was demonstrated that the amount of statistical dispersion is correlated with both the number of pairs involved in the scheme and the noise ratio. For minor random noise, it appears that one can more accurately determine the coefficients of the Hamiltonian by a lower number of pairs. It stems from the fact that for $12$ pairs of states, we obtain a $12\times 3$ matrix $\pmb{P} (\sigma)$ that is used to estimate a $3-$element vector with the coefficients of the Hamiltonian. It implies that the error of reconstruction is magnified due to the numerical estimation by the LS method.

\begin{figure}[h]
	\centering
		\centered{\includegraphics[width=0.95\columnwidth]{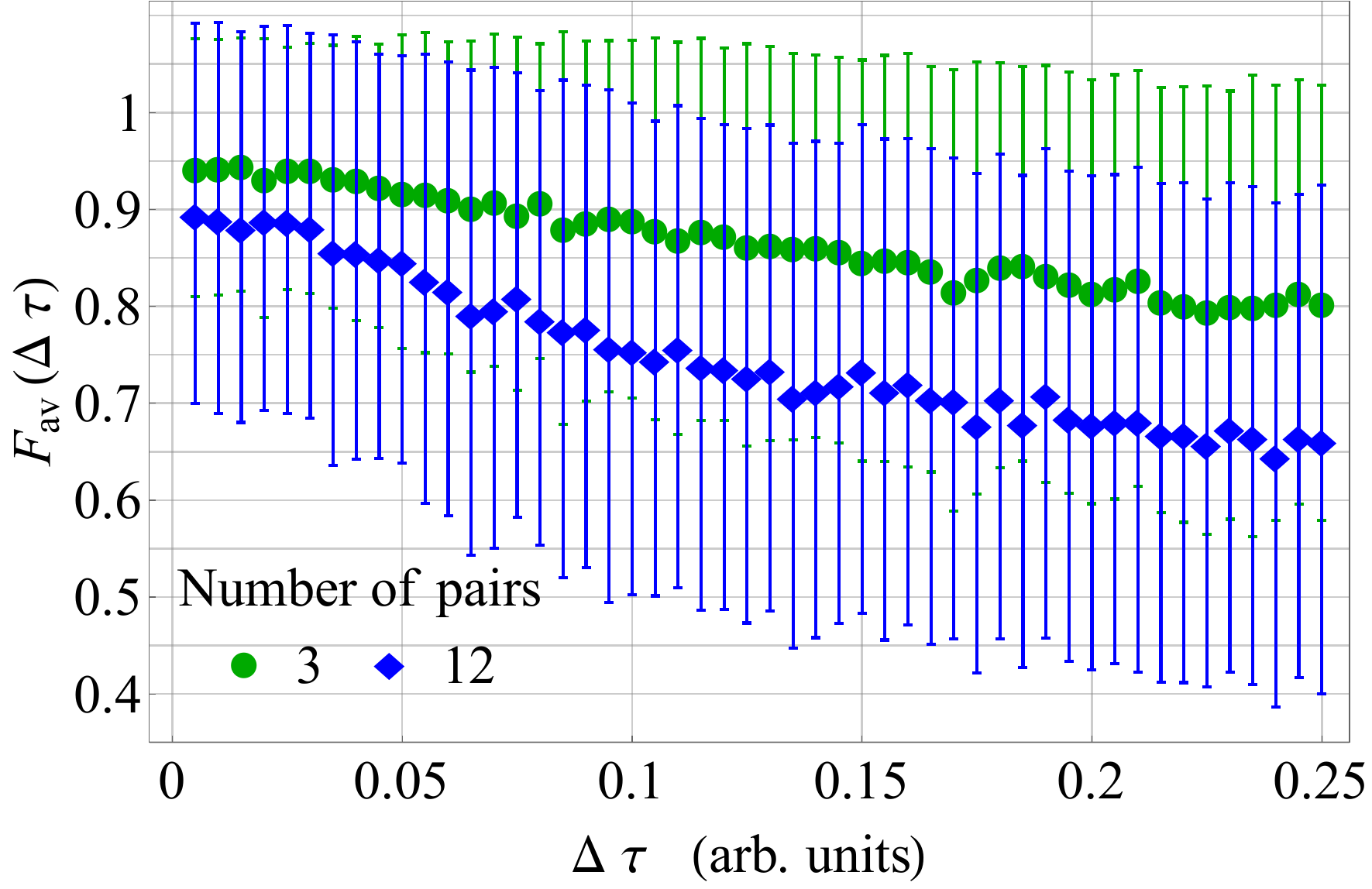}}
	\caption{Plots of $\mathcal{F}_{av} (\Delta \tau)$ for a sample of $100$ two-qubit Hamiltonians within the transverse-field Ising model. Error bars correspond to one SD. Fixed values: $\sigma = \pi/90$ and $T=1$  (arb. units).}
	\label{hamiltonian5}
\end{figure}

Next, we can test the performance of the quantum quench protocol subject to both unitary perturbations and time uncertainty. We fix one parameter, i.e. $\sigma = \pi/90$, and $\Delta \tau$ is treated as an independent variable. The value of $\sigma$ was selected in such a way that the random rotations are influencing the estimation, but they are not detrimental. Therefore, we can track the quality of reconstruction versus the amount of time uncertainty. Since the time interval, $[0, T]$, can be selected liberally provided it allows of sufficient state change due to unitary evolution, we put $T=1$ (arb. units). Henceforth, the unit of time-related parameters (i.e., $T$ and $\Delta \tau$) can be omitted since it is fixed arbitrarily. In \figref{hamiltonian5}, one finds the results for two numbers of pairs involved in the scheme.

The results demonstrate that the quality of Hamiltonian estimation gradually declines as we increase the time uncertainty. Interestingly, the plot for $12$ pairs of initial and final states lies below the graph corresponding to three pairs. For each value of $\Delta \tau$, not only is the average fidelity lower with $12$ pairs, but also SD is greater, which means that the sample features more variance. One should conclude that the scenario with $12$ pairs is more prone to random errors, which should be understood as a combined effect of the uncertainty introduced into the framework and numerical inaccuracies related to Hamiltonian estimation by the LS method.

\section{Hamiltonian tomography of three-qubit systems}\label{results3}

In this part, we implement the quantum quench protocol to determine the scalar couplings of a three-spin quantum system. Following \cite{Chen2021}, we adopt the transverse radio-frequency (rf) field as the standard of calibration, which leads to the Hamiltonian of the system in the form:
\begin{equation}\label{r9}
\begin{aligned}
{}&\mathcal{H}_3 = \alpha_1 \sigma_1  \otimes \mathbb{I}_2 \otimes \mathbb{I}_2  + \alpha_2 \mathbb{I}_2 \otimes  \sigma_1  \otimes \mathbb{I}_2 + \alpha_3 \mathbb{I}_2 \otimes  \mathbb{I}_2 \otimes  \sigma_1 +\\& +\beta_{12} \sigma_3 \otimes \sigma_3\otimes \mathbb{I}_2 + \beta_{23} \mathbb{I}_2 \otimes \sigma_3 \otimes \sigma_3+ \beta_{13} \sigma_3 \otimes \mathbb{I}_2 \otimes \sigma_3 ,
\end{aligned}
\end{equation}
where $\alpha_k = \pi \,\omega^k_{rt}$ with $\omega^k_{rt}$ standing for the rotation frequency of the $k-$th qubit driven by the rt field, and $\beta_{jk} = \pi/2 \,J_{jk}$ with $J_{jk}$ representing the scalar coupling constants between spins $j$ and $k$. Six parameters need to be determined to acquire complete knowledge about the Hamiltonian \eqref{r9}.

\begin{figure}[h]
	\centering
		\centered{\includegraphics[width=0.95\columnwidth]{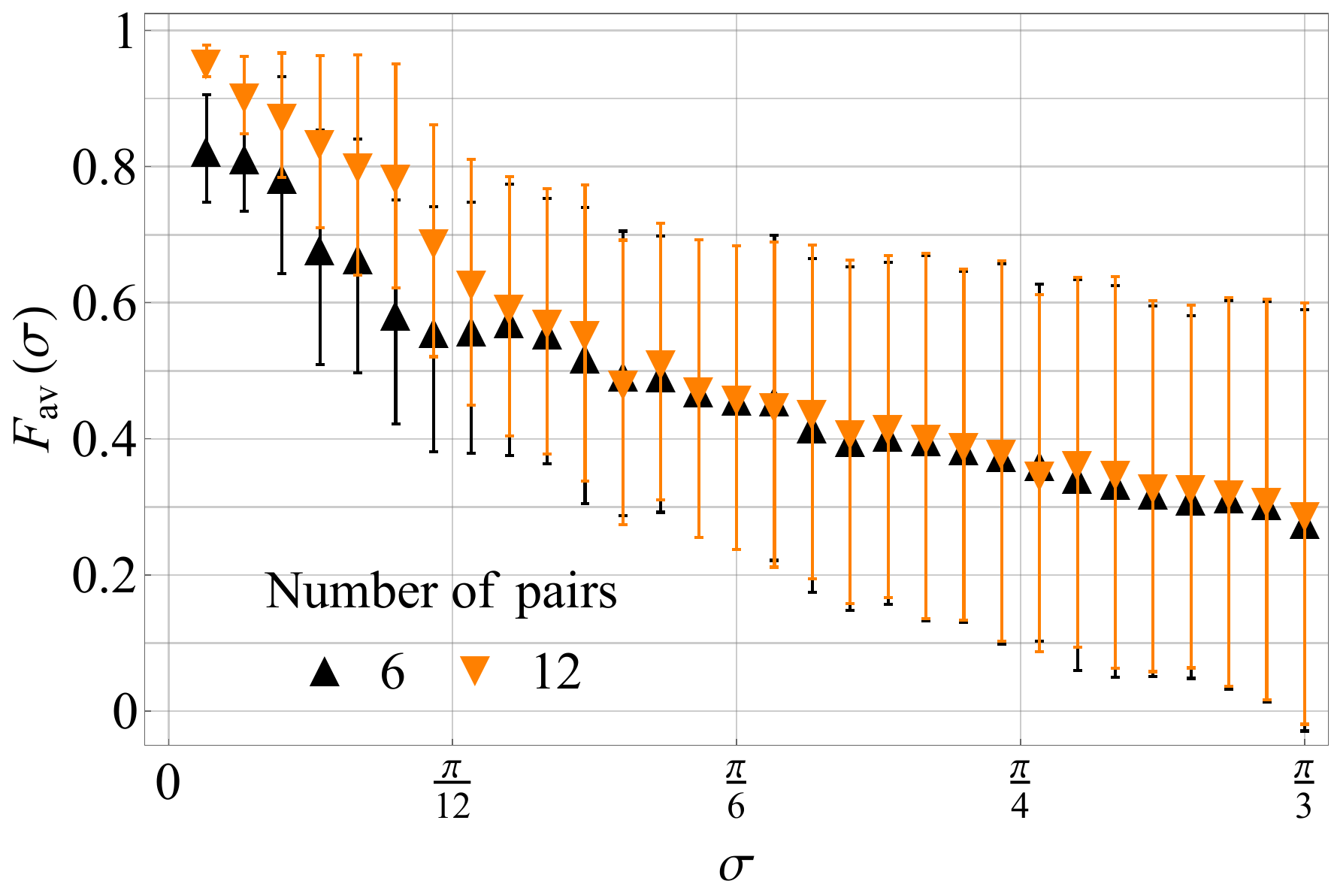}}
	\caption{Plots of $\mathcal{F}_{av} (\sigma)$ for a sample of $25$ three-qubit Hamiltonians in the form \eqref{r9}. Error bars correspond to one SD.}
	\label{hamiltonian6}
\end{figure}

The computations require more processing power due to the dimension of the Hilbert space associated with three-qubit systems. For this reason, we consider only two numbers of pairs of quantum states: $6$ and $12$. Additionally, in these simulations, the size of the sample is reduced compared with the above examples. We select a sample of $25$ random Hamiltonians in the form \eqref{r9}, for which noisy data is generated with three-qubit perturbation matrices \eqref{f61}. Nevertheless, the quantum quench protocol proved to be an efficient method for operating with three spins. In \figref{hamiltonian6}, we present the plots of $\mathcal{F}_{av} (\sigma)$, assuming no time uncertainty ($\Delta \tau=0$).

One can notice that for $\sigma< \pi/10$ the scenario that involves $12$ pairs has an advantage over the other approach. Then, for higher amounts of noise, both plots converge. To better illustrate the difference, we present specific values of the average fidelity and corresponding SDs in \tabref{table2}.

\begin{table}[h]
\setlength{\tabcolsep}{10.5pt} 
\renewcommand{\arraystretch}{2.45}
	\begin{tabular}{|c|c|c|c|c|}
\hline
\multirow{2}{*}{
				\backslashbox[15 mm]{$\sigma$}{$r$}} & \multicolumn{2}{c|}{$6$} & \multicolumn{2}{c|}{$12$}\\ \cline{2-5} 
	 	&  $\mathcal{F}_{av}$ & SD & $\mathcal{F}_{av}$ & SD \\ \hline
$\frac{\pi}{90}$ &$0.83$ & $0.08$  &$0.956$ & $0.024$    \\ \hline
$\frac{\pi}{45}$ &$0.82$ & $0.09$ &$0.91$ & $0.06$  \\ \hline
$\frac{\pi}{30}$ &$0.79$ & $0.15$ &$0.88$ & $0.09$   \\ \hline
$\frac{2 \pi}{45}$ &$0.68$ & $0.18$ &$0.84$ & $0.13$  \\ \hline
$\frac{\pi}{18}$ &$0.67$ & $0.18$ &$0.81$ & $0.17$  \\ \hline
$\frac{\pi}{15}$ &$0.59$ & $0.17$ &$0.79$ & $0.17$   \\ \hline
$\frac{7 \pi}{90}$ &$0.56$ & $0.18$ &$0.69$ & $0.17$  \\ \hline
$\frac{4 \pi}{45}$ &$0.56$ & $0.19$ &$0.63$ & $0.18$   \\ \hline
$\frac{\pi}{10}$ &$0.57$ & $0.19$ &$0.59$ & $0.19$  \\ \hline
	\end{tabular}
	\caption{Average fidelity $\mathcal{F}_{av}$ and SD for a sample of $25$ three-qubit Hamiltonians.}
	\label{table2}
\end{table}

The results allow one to conclude that random unitary rotations have a detrimental impact on the quality of Hamiltonian tomography. Nevertheless, for minor values of $\sigma$, the quantum quench protocol has proved to be efficient.

\begin{figure}[h]
	\centering
		\centered{\includegraphics[width=0.95\columnwidth]{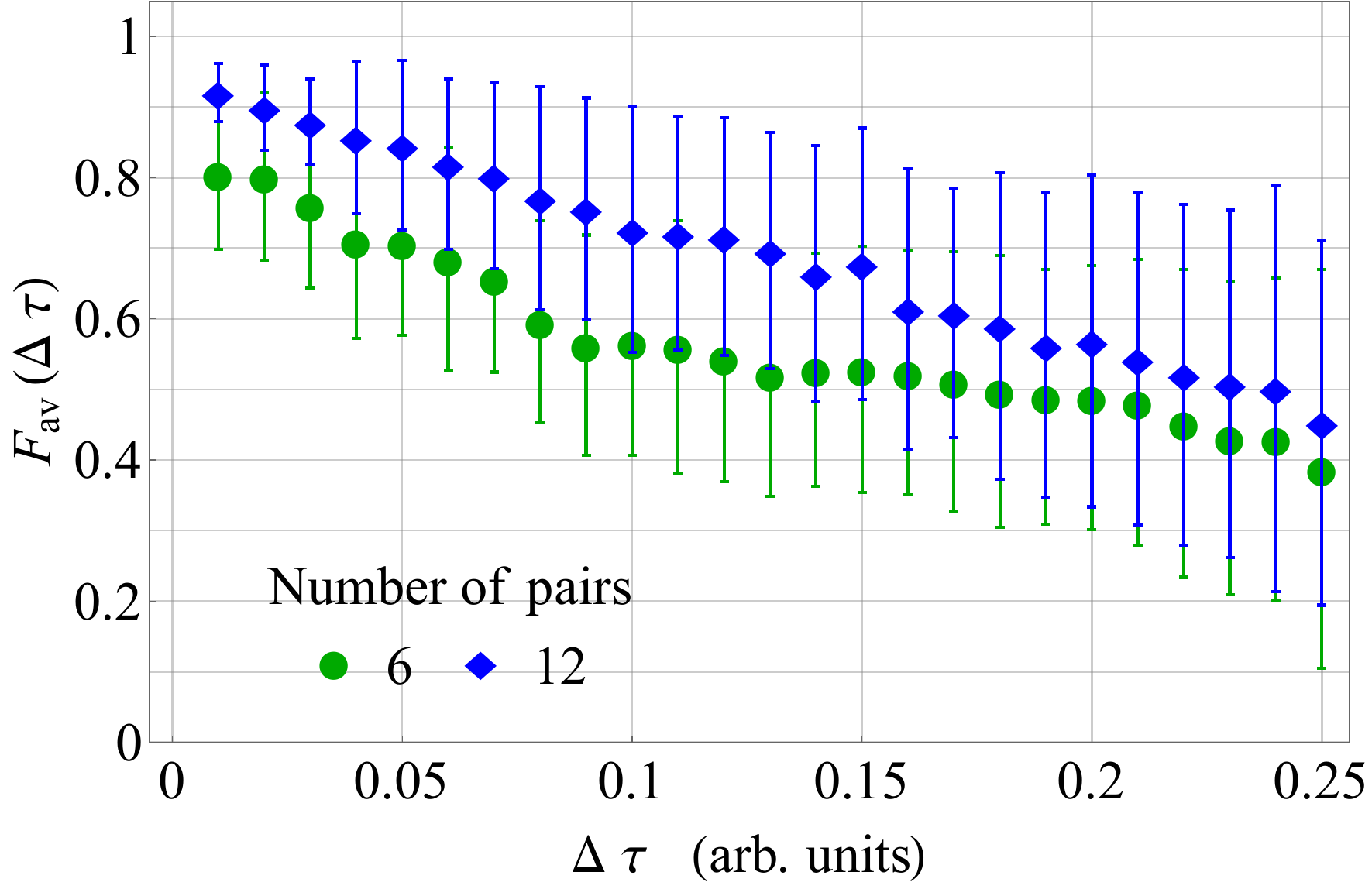}}
	\caption{Plots of $\mathcal{F}_{av} (\Delta \tau)$ for a sample of $25$ three-qubit Hamiltonians. Error bars correspond to one SD. Fixed values: $\sigma = \pi/90$ and $T=1$  (arb. units).}
	\label{hamiltonian7}
\end{figure}

Secondly, we investigate the robustness of the protocol by imposing time uncertainty on the moments of measurement. We keep random unitary rotations with $\sigma = \pi/90$, and randomly select the moment of measurement for each pair from a normal distribution characterized by a standard deviation $\Delta \tau$. In \figref{hamiltonian7}, the plots of $\mathcal{F}_{av} (\Delta \tau)$ are presented. As we increase the value of $\Delta \tau$, both plots converge while the corresponding SD grows. For a minor extent of noise, we obtain a satisfying quality of QHT. In particular, $\mathcal{F}_{av} ( 0.01) = 0.92$ with $SD=0.04$ for $12$ pairs. This outcome proves that the quantum quench protocol provides decent results even in the presence of two kinds of random noise. Since the statistical dispersion increases as we add more time uncertainty, the task of the experimenter is to constrain the impact of such errors by implementing state-of-the-art devices that are accurately calibrated.

\section{Discussion and outlook}\label{discussion}

In the article, we introduced a framework for numerical simulations of QHT with random noise. In our model, the noise was introduced in two ways: by random unitary rotations that distort the actual measurement operators and by time uncertainty related to the moments of measurement. Two parameters, $\sigma$ and $\Delta \tau$, were used to quantify the noise. The figures of merit defined to measure the performance of the framework were presented on graphs versus the amount of noise. Each time, the simulations were performed for a wide range of parameters to test the theoretical robustness of the framework under conditions of heave noise.

The framework was tested on three types of one-qubit Hamiltonians, and the results appear very consistent. We proved that if the random noise is not severe ($\sigma < \pi/10$), we can efficiently perform QHT with three pairs of initial and final states. Increasing the number of pairs leads to a growth of statistical dispersion, which makes the results for the sample more scattered. For a heavier amount of random noise, we cannot precisely determine the Hamiltonian, although the schemes based on more pairs provided better average fidelity. One can agree that in experiments, we are able to reduce the impact of random errors connected uncertainties in apparatus settings. Therefore, the condition for precise Hamiltonian estimation appears to lie within the reach of current technology.

The main results of the article involve QHT of multi-qubit systems. First, we implemented the quantum quench protocol to a two-qubit Hamiltonian within the transverse-field Ising model.  Both sources of experimental errors were considered. For any number of pairs, the function $\mathcal{F}_{av} (\sigma)$ is convex whereas $\mathcal{F}_{av} (\Delta \tau)$ is approximately linear. The results proved that under conditions of random time uncertainty, one can always determine the Hamiltonian more accurately with a lower number of pairs. As for random unitary rotations, the difference is noticeable only for minor values of noise. The results come from the fact that the errors associated with multiple pairs get magnified in the estimation algorithm that is based on the LS method. For a high noise rate, individual measurement results deliver conflicting information about the Hamiltonian in question, and increasing the number of pairs cannot improve the efficiency. A possible solution to this problem may involve increasing the number of iterations or applying machine learning methods to find an optimal fit for the noisy data, see for example Ref.~\cite{Che2021}.

Furthermore, we introduced results devoted to the Hamiltonian estimation of three-qubits systems. In this case, it was demonstrated that increasing the number of pairs of states involved in the framework results in a better quality of Hamiltonian tomography, which was particularly evident for moderate amounts of noise. This conclusion is in agreement with experimental research devoted to three-body Hamiltonians \cite{Chen2021}. Even if we assume two sources of random errors, the quantum quench protocol guarantees a reliable three-qubit Hamiltonian reconstruction up to a certain level of noise.

Also, it is worth highlighting that the geometric properties of the initial states were not considered a variable that could influence the efficiency of the framework. According to \cite{Li2020}, randomly selected initial states perform poorly in the quantum quench protocol. Therefore, following this point of view, the states in the initial ensemble for each type of Hamiltonian were selected arbitrarily, bearing in mind that the geometric distance between any two of them should be sufficient to gain independent information about the unknown Hamiltonian. However, it requires more research to properly understand the relation between the geometry of input states and the quality of Hamiltonian estimation. Saying that initial states should be distinct enough is not a comprehensive explanation. Thus, a natural question for further research relates to the interdependence between the quality of QHT and geometry of initial states.

Finally, it should be noticed that the quantum quench protocol suffers not only from the impact of random noise, but it is inherently restricted to determining just the coefficients that appear in the Hamiltonian decomposition. We must a priori know the basis that has to consist of operators representing measurable quantities. In spite of providing a theoretical approach, the present article focused on examples of Hamiltonians that are introduced in bases defined by operators traditionally considered measurable.

In the future, the framework will be extended and applied to other Hamiltonians of multi-level systems because the efficiency of the method appears to depend on the algebraic properties of $\mathcal{H}$. Additionally, different kinds of experimental noise will be mathematically modeled and incorporated into the scheme to test the performance of the framework. Furthermore, the quantum quench protocol appears to be an advantageous technique that can be implemented to develop the verification of quantum devices \cite{Carrasco2021} or entanglement Hamiltonian tomography \cite{Kokail2021}. Last but not least, the Hamiltonian tomography will be investigated in relation to the security of blind quantum computing for a fixed Hamiltonian \cite{Qiang2017,Quan2021}, including a representation as a linear combination of unitary operators \cite{Long2006}. In particular, high dimensional cases will be studied since qudits have significant advantages, including higher information capacity, improved security against eavesdroppers, and better resilience to noise \cite{Romero2019}.

\section*{Acknowledgments}

The research was supported by the National Science Centre in Poland, grant No. 2020/39/I/ST2/02922.

\end{document}